# Propagation dynamics of radially polarized symmetric Airy beams in the fractional Schrödinger equation

Shangling He[1], Boris A. Malomed[2,3], Dumitru Mihalache[4], Xi Peng[5], Yingji He[5], and Dongmei Deng[1]*

[1]*Guangdong Provincial Key Laboratory of Nanophotonic Functional Materials and Devices, South China Normal University, Guangzhou 510631, China*

[2] *Department of Physical Electronics, School of Electrical Engineering, Faculty of Engineering, and Center for Light-Matter Interaction, Tel Aviv University, Tel Aviv 69978, Israel*

[3]*Instituto de Alta Investigación, Universidad de Tarapacá, Casilla 7D, Arica, Chile*

[4]*Horia Hulubei National Institute for Physics and Nuclear Engineering, P.O. Box MG-6, RO-077125, Bucharest-Magurele, Romania*

[5]*School of Photoelectric Engineering, Guangdong Polytechnic Normal University, Guangzhou 510665, China*

*Corresponding author: dmdeng@263.net

## Abstract

We analyze the propagation dynamics of radially polarized symmetric Airy beams (R-SABs) in a (2+1)-dimensional optical system with fractional diffraction, modeled by the fractional Schrödinger equation (FSE) characterized by the Lévy index, $\alpha$. The autofocusing effect featured by such beams becomes stronger, while the focal length becomes shorter, with the increase of $\alpha$. The effect of the intrinsic vorticity on the autofocusing dynamics of the beams is considered too. Then, the ability of R-SABs to capture nano-particles by means of radiation forces is explored, and multiple capture positions emerging in the course of the propagation are identified. Finally, we find that the propagation of the vortical R-SABs with an off-axis shift leads to rupture of the ring-shaped pattern of the power-density distribution.

## 1. Introduction

In 2007, finite-energy (truncated) Airy beams were proposed theoretically and experimentally by Siviloglou *et al*. [1,2]. These and subsequent works had revealed novel features of Airy modes, such as self-acceleration, self-healing, and self-autofocusing [3-7]. Then, the Airy waves were extended to cylindrical coordinates, leading to the creation of beams with circular symmetry and the Airy radial profile. Such circular Airy beams feature a regime of abrupt autofocusing [8-10], which can be used, in particular, for trapping small particles [9].

In 2014, symmetric Airy beams (SABs) were proposed by Vaveliuk *et al*. They arise from the finite-energy Airy beam, changing the odd parity into even in the spectral cubic phase, i.e., replacing factor $k^3$ in the phase of the Fourier transform of the truncated Airy wave, with wavenumber $k$, by $|k|^3$ [11-13]. The spectral phase of the two-dimensional (2D) SAB features a quadratic pattern. SABs demonstrate strong autofocusing in the course of the propagation, which offers additional possibilities for applications to optical trapping and manipulations [14-19].

While the usual Airy beams carry linear polarization, radially polarized symmetric Airy beam (R-SAB) were proposed by Xu *et al*. [20]. They also feature square-shaped phase patterns and tighter self-autofocusing.

In another research area, fractional Schrödinger equations (FSEs), involving spatial derivatives of non-integer orders, were introduced by Laskin as quantum-mechanical equations for the wave function of particles moving by Lévy flights [21], the main characteristic of the FSE being the respective Lévy index (LI). Since then, FSEs became a subject of extensive studies [22-27]. A realization of FSE in terms of classical optics, based on transverse light dynamics in aspherical optical cavities, was proposed by Longhi [28]. The propagation of beams, in the form of as 1D and 2D input chirped Gaussians and super-Gaussians, in FSE optical system with effective external potentials was elaborated too [29-31]. In particular, LI may be used to control the period of Rabi oscillations and resonant-conversion distances in the FSE [32]. Various aspects of soliton dynamics in nonlinear FSE models have also been theoretically investigated, including emission of radiation [33], spontaneous symmetry breaking [34], vortex solitons [35], necklace-shaped patterns [36], nonlinear lattices [37], and dissipative solitons in fractional complex Ginzburg-Landau equation [38]. In all these contexts, LI is, naturally, the most important control factor.

A different context in which fractional spatial derivatives naturally appear is the continuum limit of dynamical lattices with long-range inter-site interactions [39]. As a result, one derives a Klein-Gordon equation with the spatial dispersion represented by fractional derivatives. Then, the usual long-wave expansion reduces the latter equation to the FSE form.

The autofocusing of Airy beams in the framework of the linear FSE was recently studied in works [40] and [41]. However, the propagation of a radially polarized beams (including ones shifted off the center) governed by FSE, which is the subject of the present work, was not previously addressed. We also consider possibilities to use the self-acceleration and autofocusing effects, featured by R-SABs in the framework FSE, for the capture of particles. In this Letter, the FSE is introduced in Section 2.

Basic numerical results of the analysis are summarized in Section 3, which is followed by consideration of an especially interesting issue, *viz.*, the off-axis propagation of the SAB, in Section 4. The paper is concluded by Section 5.

2. **The theoretical model**

We consider the propagation dynamics of R-SABs modeled by the vector (two-component) FSE [22,23,42],

$$i\frac{\partial \vec{u}}{\partial z} - \frac{1}{2kw_0^{2-\alpha}}\left(-\frac{\partial^2}{\partial x^2} - \frac{\partial^2}{\partial y^2}\right)^{\alpha/2}\vec{u} = 0, \qquad (1)$$

where $z$ is the longitudinal propagation distance, $x$ and $y$ are scaled transverse coordinates, $k = 2\pi/\lambda$ is the carrier wavenumber, $\lambda$ is the wavelength of the incident light, $w_0$ is a scaling factor, $\vec{u}$ is the vector field of the optical wave, and $\alpha$ is LI ($1 < \alpha \leq 2$). The fractional-diffraction operator in Eq. (1) is defined by the known integral expression [16-17, 36],

$$\frac{1}{2w_0^{2-\alpha}}\left(-\frac{\partial^2}{\partial x^2} - \frac{\partial^2}{\partial y^2}\right)^{\alpha/2} u(x,y,z)$$
$$= \frac{1}{8\pi^2 w_0^{2-\alpha}} \iint dk_x dk_y \exp(ik_x x + ik_y y)(k_x^2 + k_y^2)^{\alpha/2} \hat{u}(k_x, k_y, z), \qquad (2)$$

where the two-dimensional Fourier transform of the field, which is a function of spatial frequencies $k_x$ and $k_y$, is

$$\hat{u}(k_x, k_y, z) = \iint dxdy \exp(-ik_x x - ik_y y) u(x,y,z).$$

For $\alpha = 2$, Eq. (1) reduces to the usual two-dimensional linear Schrödinger equation.

Starting from the expression for the 2D symmetric Airy wave, superimposed by the spiral phase with winding number (vorticity) *l*, written in the Cartesian coordinates, the solution is looked for as

$$u_x = \frac{x}{x_0} u_1\left(\frac{x}{x_0}, \frac{y}{y_0}\right) \Pi_{j=1}^{N} \frac{\left[(x-x_j)+i(y-y_j)\right]^l}{\left[(x-x_j)^2+(y-y_j)^2\right]^{l/2}}, \quad (3)$$

$$u_y = \frac{y}{y_0} u_1\left(\frac{x}{x_0}, \frac{y}{y_0}\right) \Pi_{j=1}^{N} \frac{\left[(x-x_j)+i(y-y_j)\right]^l}{\left[(x-x_j)^2+(y-y_j)^2\right]^{l/2}}, \quad (4)$$

where $u_{x,y}$ are complex amplitudes of the two components of the vector field. Transverse length scales $x_0$ and $y_0$ (actually, solutions are presented below for $x_0 = y_0$) determine the Rayleigh (diffraction) distance for the beams, $Z_R = kx_0^2$, and $(x_j, y_j)$ are coordinates of the $j$-th vortex phase dislocation, $N$ being the number of vortices. Further, $u_1(x,y)$ represents the 2D symmetric Airy wave [43]:

$$u_1(x,y,z=0) = u_0(s_x) u_0(s_y)$$
$$= \frac{1}{4}\{\exp(-\aleph s_x)[\mathrm{Ai}(-s_x)+i\mathrm{Gi}(-s_x)] + \exp(\aleph s_x)[\mathrm{Ai}(s_x)+i\mathrm{Gi}(s_x)]\}$$
$$\times \{\exp(-\aleph s_y)[\mathrm{Ai}(-s_y)+i\mathrm{Gi}(-s_y)] + \exp(\aleph s_y)[\mathrm{Ai}(s_y)+i\mathrm{Gi}(s_y)]\}, \quad (5)$$

where $u_0(s_{x,y})$, with $s_x \equiv x/x_0$, $s_y \equiv y/y_0$, is the known expression for the 1D SAB in the free space, $\mathrm{Ai}(s_{x,y})$ is the Airy function, $\aleph$ is the truncation factor, and $\mathrm{Gi}(s_{x,y}) \equiv \frac{1}{\pi}\int_0^{\infty} \sin\left(\frac{1}{3}t^3 + s_{x,y}t\right) dt$. Parameters chosen for the subsequent analysis correspond to $w_0 = x_0 = y_0 = 1$mm in physical units, and the truncation factor in Eq. (5) is fixed as $\aleph = 0.2$.

3. **Numerical results**

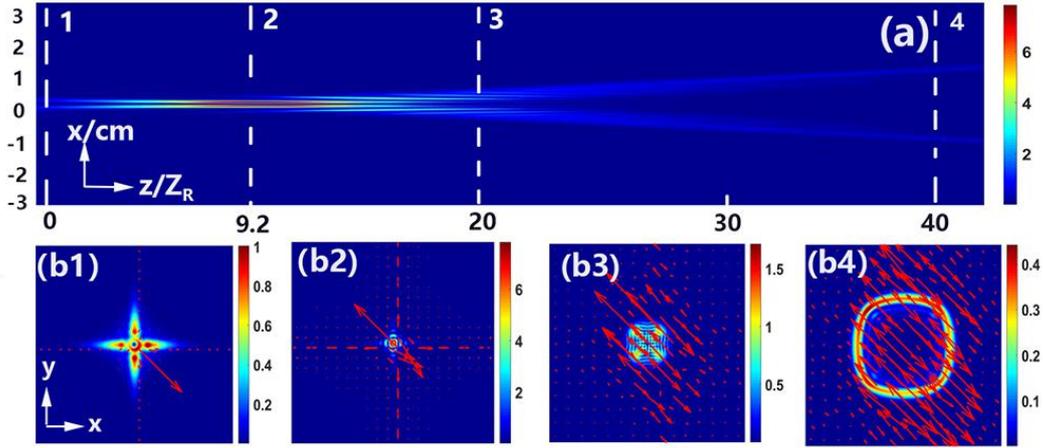

Fig. 1. (color online) Propagation of the R-SAB input (with $l = 0$, i.e., no vorticity, in Eqs. (3) and (4)), as produced by numerical simulations of the FSE (1) with LI $\alpha =1.4$. (a) The side view. (b1)-(b4) Snapshots of transverse intensity patterns (the background) and polarization (arrows) in planes marked as 1-4 in (a), which correspond, severally, to propagation distances $z = 0, 9.2Z_R, 20Z_R,$ and $40Z_R$.

The propagation of the R-SAB, defined as Eqs. (3) - (5) with $l = 0$, governed by Eq. (1) with LI $\alpha =1.4$ is shown in Fig. 1. This value is chosen here, as well as in Figs. 5 – 7 below, as it helps to report generic results obtained for LI which is sufficiently different form the one, $\alpha = 2$, which corresponds to the usual paraxial diffraction. It is seen in panel (a) that the beam sharply autofocuses into the tightest doughnut-shaped spot after a certain propagation distance. Compared to the R-SAB propagation in the usual (2+1)-dimensional setting [21], the focusing length becomes longer (the first focus point is located at distance $z = 9.2Z_R$), the focusing area becomes larger, and the diffraction distance too is longer in the case of the fractional diffraction.

As shown in Fig. 1(b1), the initial R-SAB concentrates its power, mainly, in four side lobes. In the course of the propagation, the four lobes merge into the single doughnut-shaped spot in Fig. 1(b2). Past the focal plane, the beam gradually expands outward and forms a hollow square structure in Fig. 1(b3), with the power chiefly concentrated at four corners of the square. The presence of the hollow spot at the center of the beam offers an obvious advantage for using it as a tool capturing small particles, which may be placed in the hollow area, as discussed below. As the propagation distance keeps growing, the transverse intensity distribution gradually changes from the hollow square into a larger hollow rounded pattern in Fig. 1(b4). The polarization of the evolving structure is mainly directed in positive and negative directions, at angle $\approx (3/4)\pi$ with the $x$-axis. In this connection, it is relevant to mention that the equations for $x$- and $y$-components of the vector field in Eq. (1) are not really coupled, hence the solution can be split into scalar components. Nevertheless, the representation in the vector form is useful, as it helps to display the result in a clear form, especially in the case of R-SABs carrying intrinsic vorticity, as shown below.

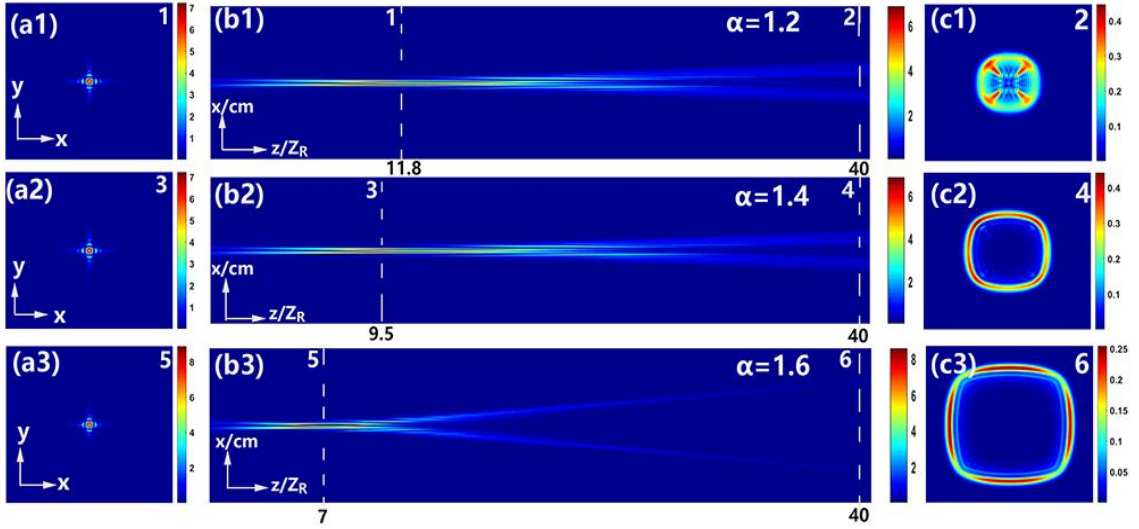

Fig. 2. (color online) The propagation of the R-SABs (with $l = 0$) governed by Eq. (1) with different LIs, $\alpha = 1.2, 1.4,$ and $1.6$. Panels (a1)-(a3) and (c1)-(c3) are snapshots of transverse intensity patterns in the planes marked in panels (b1)-(b3), which show side views of the propagation. Other parameters are the same as in Fig. 1.

Next, we address characteristics of the autofocusing of the R-SABs governed by the FSE (1) with different LIs. The value of the LI, $\alpha$, obviously affects the focal length and focal intensity in Figs. 2(b1)-2(b3). Before autofocusing occurs, the beams converge slowly. Then the R-SAB sharply autofocuses at distances $z = 11.8Z_R, 9.5Z_R,$ and $7Z_R$, which correspond to $\alpha = 1.2, 1.4,$ and $1.6,$ respectively. Thus, the focal length decreases and focal intensity increases with the increase of $\alpha$. Beyond the autofocusing point, the diffraction gradually enhances with the increase of $\alpha$. Accordingly, as shown in Figs. 2(c1)-2(c3), the transverse intensity of the beams builds a hollow rounded pattern for a sufficiently large propagation length, such as the on displayed in Fig. 2(c3) at $z = 40Z_R$. In particular, the size of the rounded pattern is much larger for $\alpha = 1.6$ than for $\alpha = 1.2$, with the same propagation distance.

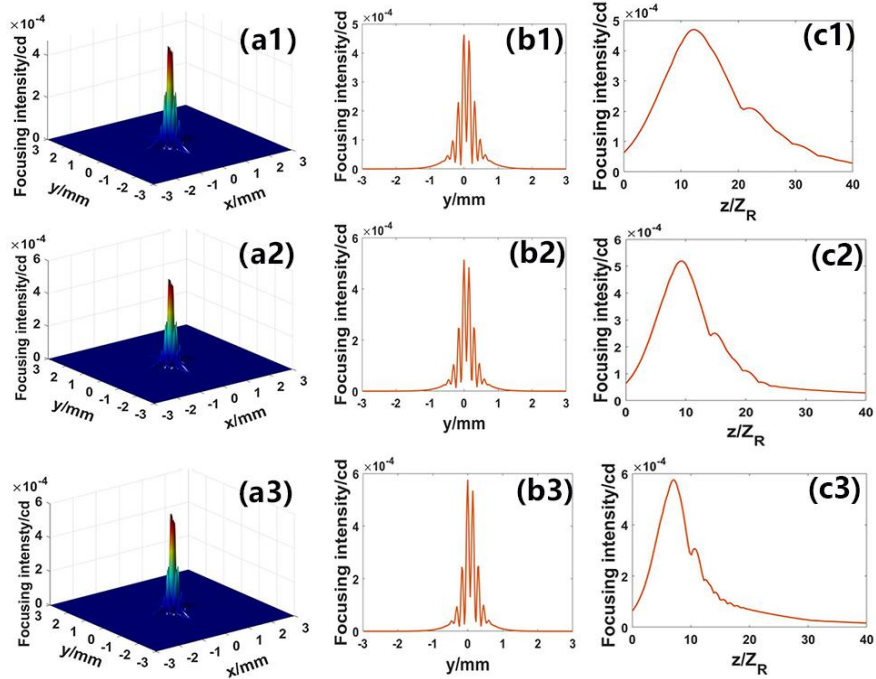

Fig. 3. (color online) (a1)-(a3) Transverse intensity profiles of the R-SABs in the focal plane. (b1)-(b3) The intensity distributions in the cross section $x = 0$. (c1)-(c3) The peak intensity vs. the propagation distance. The parameters are $\alpha = 1.2$ in (a1)-(c1), $\alpha = 1.4$ in (a2)-(c2), and $\alpha = 1.6$ in (a3)-(c3).

Figures 3(a1)-3(a3) show the transverse intensity profiles of the R-SABs in the focal plane. The intensities in these planes and the evolution of the peak intensity for $\alpha = 1.2$, $\alpha = 1.4$ and $\alpha = 1.6$ are shown in Figs. 3 (b1)-3(b3) and 3(c1)-3(c3), respectively. In contrast to the ordinary Airy beams [44], the radius of the ring of the beams in the focal plane decreases with the increase of $\alpha$, and the focal spot is hollowed at its center. The reason for this difference is that the R-SABs contain two autofocusing components. In Figs. 3(c1) and 3(c3), close to the focal point, the power of the R-SABs is concentrated in a small area and the focusing intensity sharply increases. Beyond the autofocusing point, the decrease of the peak intensity is not monotonous, showing oscillations related to the formation of secondary Airy rings.

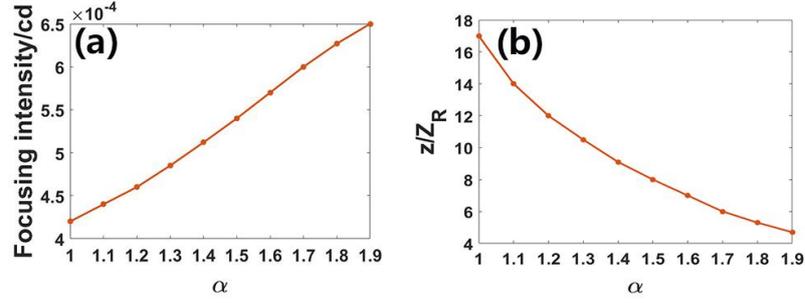

Fig. 4. (color online) Panels (a) and (b): the focal intensity and length of the R-SAB vs. the LI (Lévy index), $\alpha$. Other parameters are the same as in Fig. 1.

The focal intensity and the respective focal length of the R-SABs are shown, as functions of the LI, in Figs. 4(a) and 4(b), respectively. It is seen that the focal intensity and distance monotonously increases and decreases, respectively, with the increase of the LI. Thus, larger LI naturally leads to faster autofocusing, as the diffraction term in Eq. (1) becomes stronger with the increase of $\alpha$.

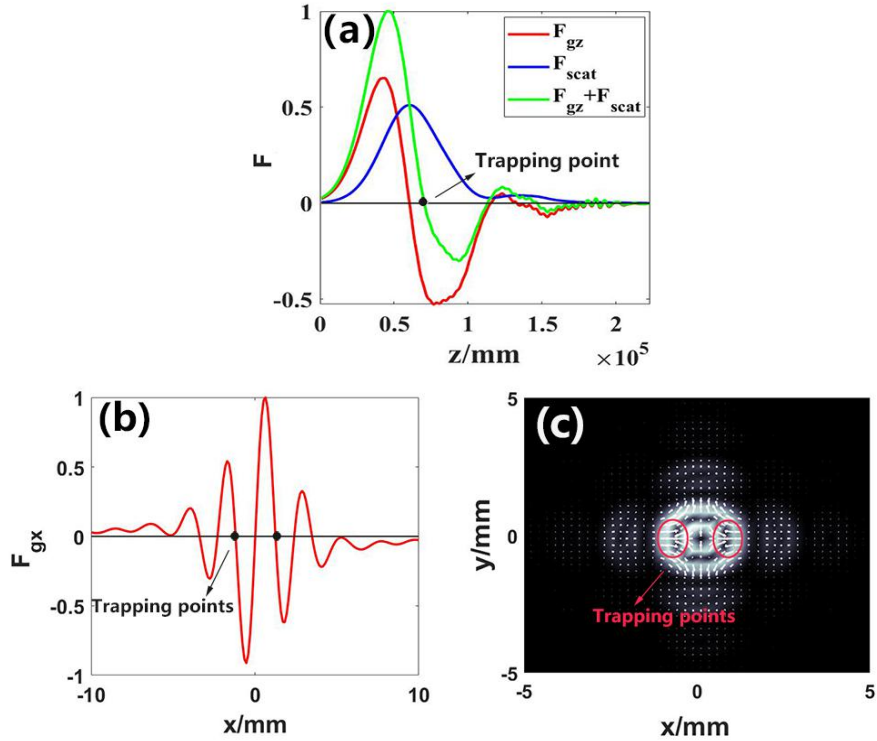

Fig. 5. (color online) (a) The distribution of the radiation force of the R-SABs along the $z$- axis, in arbitrary units; (b) the distribution of the gradient force along the $x$-axis; (c) the distribution of the gradient force in the focal plane. The forces are calculated as per Eqs. (6) and (7). All the parameters are the same as in Fig. 1, while the size of the probe particle, onto which the forces are acting, is taken as $\mu_0$=0.15 nm, in physical units.

Radiation forces acting in the R-SAB field patterns, which determine the trapping ability of the R-SAB, contain two components, the gradient and scattering ones. These forces are defined as [45]

$$\overrightarrow{F_{grad}}(x,y,z) = \frac{2\pi\mu_0^3 n_2}{c}\left(\frac{m^2-1}{m^2+2}\right)\nabla I(x,y,z), \qquad (6)$$

$$\overrightarrow{F_{scatt}}(x,y,z) = \frac{8\pi k^4 \mu_0^6 n_2}{3c}\left(\frac{m^2-1}{m^2+2}\right)I(x,y,z)\overrightarrow{e}_z, \qquad (7)$$

where $m = n_1/n_2$ is the relative refractive index of the probe nano-particle (here, we assume $n_1 = 1.5$ and $n_2 = 1$ for the medium), $\mu_0$ is the particle's radius, and the intensity is $I(x,y,z) = \frac{1}{2}cn_2\varepsilon_0|\vec{u}(x,y,z)|^2 \equiv \frac{1}{2}cn_2\varepsilon_0[|u_x(x,y,z)|^2 + |u_y(x,y,z)|^2]$, where $\vec{u}(x,y,z)$ is the vector field governed by Eq. (1). The calculation, based on Eqs. (6) and (7), yields the fields of the longitudinal and transverse radiation forces acting along the $z$ and $x$ axes, which are displayed, respectively, in Figs. 5(a) and 5(b). When the particle's radius $\mu_0$ is smaller, the gradient force is larger, helping to capture the particle in the longitudinal direction. Note that the position of the point at which the nano-particle can be trapped is slightly different from the focal point because of the scattering force. There are actually two trapping points, shown by black bold dots on the $x$ axis in the plot of the gradient force in Fig. 5(b). Accordingly, the two capture positions on the $x$ axis are identified in Fig. 5(c). In addition, three other trapping points are located at three hollowed spots on the $y$ axis, where the nano-particle may be trapped as well. These results suggest direct applications for optical manipulations of small particles.

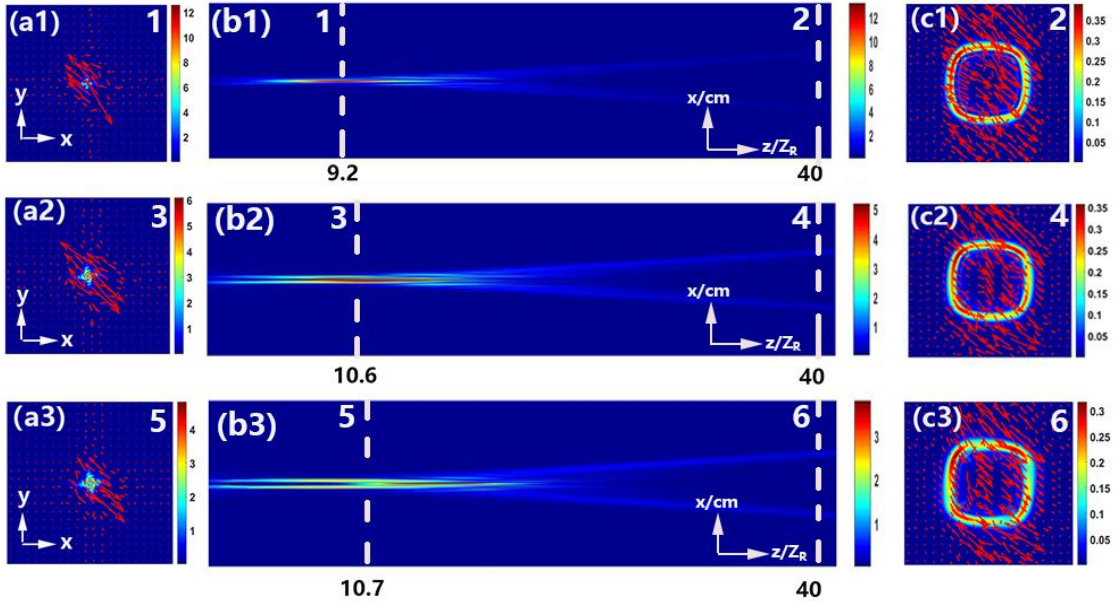

Fig. 6. (color online) Numerical illustration of the evolution of R-SABs for different values of winding number (topological charge) $l$, produced by simulations of Eq. (1) with $\alpha = 1.4$ and the input given by Eqs. (3) and (4) with $N = 1$ and $x_1 = y_1 = 0$. Panels (a1)-(a3) and (c1)-(c3) display snapshots of transverse intensity patterns and local orientation of the polarization in the planes marked by numbers 1 – 6 in panels (b1)-(b3). Panels (b1)-(b3) provide side views of the propagation with different winding numbers $l$=1, 2, and 3, respectively. Other parameters are the same as those in Fig. 1.

Next, we address effects of a nonzero winding number, $l$, in the input defined by Eqs. (3) and (4), on the evolution of R-SABs. Figure 6 demonstrates that, similar to what was shown above, in the post-autofocusing stage the R-SABs again keep their power in hollow ring-shaped patterns. A difference is that the pattern rotates in the anticlockwise direction, with an angular velocity increasing with the increase of $l$. In addition, the focal distance of the R-SABs slightly increases, while the intensity at the focal point decreases, see Figs. 6(b1)-6(b3). Moreover, the diameter of the post-autofocusing hollow ring becomes larger with the increase of $l$. Thus, we conclude that the vorticity (i.e., the beam's angular momentum) impedes the development of the autofocusing, due to the action of the "centrifugal force" created by the angular momentum.

4. **Off-axis propagation dynamics of autofocusing R-SABs**

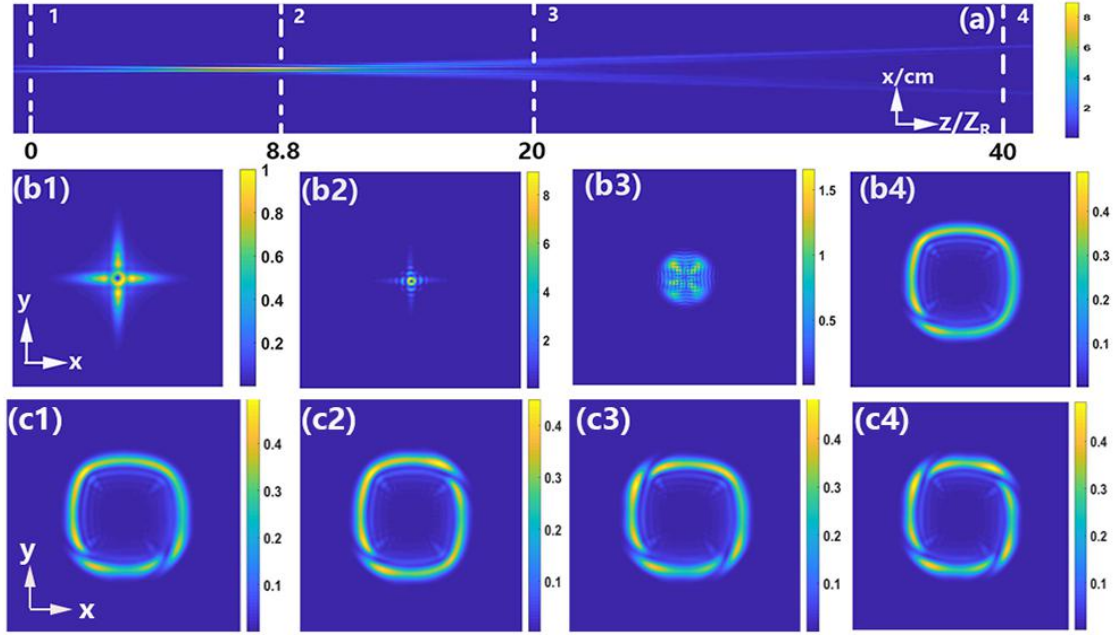

Fig. 7. (color online) (a) The side view of the off-axis R-SAB (with $N = l = 1$) produced by the input given by Eqs. (3) and (4) with an off-axis shift, $x_1 = y_1 = 4.8$ mm, in physical units. (b1-b4): Snapshots of the transverse intensity patterns of the R-SABs in planes 1-4 marked in (a). (c1-c4): snapshots of transverse intensity patterns of R-SAB with multiple off-axis vortices ($N > 1$) produced by the input given by Eqs. (3) and (4) with $|x_j| = |y_j| = 4.8$ mm. Other parameters are the same as those in Fig. 1.

Figure 7 demonstrates the propagation of the vortical R-SAB, initiated by the input given (3) and (4) with $l = 1$, shifted off-axis by $x_1 = y_1 = 4.8$ mm (in physical units). It is seen that the propagation is similar to that in the on-axis regime, with a difference that a defect emerges in the ring-shaped pattern, see Fig. 7(b4). In other words, the off-center shift breaks the ring, making it unclosed. The reason is that power flows from the left lobe (in Fig. 7(b1)) to another one. In comparison to Fig. 1, both the focal intensity and focal distance are made slightly larger by the off-axis shift, while the contrast of the (unclosed) ring-shaped pattern becomes weaker. Figures 7(c1)-7(c4) show the transverse power distribution of R-SABs with multiple off-axis vortices. They include, severally, two vortices on one side, two vortices on the diagonal, three vortices on the diagonal, and four vortices, featuring several breakups in the ring-shaped pattern. We have found that the R-SABs can carry multiple vortices and propagate stably over a long distance (up to $z = 40Z_R$ or longer).

## 5. Conclusion

We have presented systematic results for the propagation dynamics of the R-SABs

(radially polarized symmetric Airy beams), obtained in the framework of the model of a linear optical medium based on the FSE (fractional Schrödinger equation). The propagation of the R-SABs in this setup shows autofocusing, followed by the diffractive expansion, affected by value of the LI (Lévy index) and vorticity carried by the R-SAB. The autofocusing gets stronger, while the focal length and radius of the ring pattern at the focus decrease, with the increase of LI. Further, the autofocusing effect slightly weakens with the increase of the vorticity (winding number), embedded in the beam, from 1 to 3. In addition, we study the ability of the R-SABs to trap a nano-particle by the radiation force. The results identify several capture positions appearing in course of the beam's propagation. Then, we address the off-axis propagation dynamics of the R-SAB, which demonstrates the formation of the defect in the form of the disruption in the final ring-shaped pattern of the power density. The evolution of the R-SABs in the optical system with the fractional diffraction offers better performance for some applications, such as the capture of nano-particles and design of optical tweezers.

## FUNDING INFORMATION

This work was supported by the National Natural Science Foundation of China (11775083, 11374108, 61675001, 11947103, 12004081), Science and Technology Program of Guangzhou (No. 2019050001), and Israel Science Foundation (grant No. 1286/17).

## CONFLICT OF INTEREST

We declare that we have no conflict of interest.